\documentclass{raa}            
\usepackage{graphicx,times, natbib, amsmath, hyperref, enumerate}
\def\apj{{ApJ}}                 
\def\apjl{{ApJ}}                
\def\apjs{{ApJS}}               
\def\ao{{Appl.~Opt.}}           
\def\aap{{A\&A}}                
\def\aaps{{A\&AS}}              
\def\mnras{{MNRAS}}             

\def\jcap{{JCAP}}               

\def\physrep{{Phys.~Rep.}}   

\begin{document}

\title{Influence of baryons on spatial distribution of matter: higher order correlation functions}

\volnopage{ {\bf 20xx} Vol.\ {\bf x} No. {\bf XX}, 000--000}
\setcounter{page}{1}
\author{Xiaojun Zhu \inst{1}  \and Jun Pan \inst{1,2} }
\institute{The Purple Mountain Observatory, 2 West Beijing Road, Nanjing 210008, China \\ 
\and National Astronomical Observatories, Chinese Academy of Sciences, Beijing 100012, China }
\date{Received~~2012 month day; accepted~~2012~~month day}

\abstract{Baryonic physical processes could leave non-negligible imprint on cosmic matter distribution.  
Series of high resolution simulation data sets with identical initial condition 
are employed for count-in-cell (CIC) analysis, including
one N-body pure dark matter run, one with adiabatic gas only and one with dissipative processes. 
Variances and higher order cumulants $S_n$ of dark matter and gas
are estimated. It is found that baryon physical processes mainly 
affected dark matter distribution at scales less than $1h^{-1}$Mpc. In comparison with the 
pure dark matter run, adiabatic process alone strengthens variance of dark matter 
by $\sim 10\%$ at scale $0.1h^{-1}$Mpc, while $S_n$s of dark matter
deviate only mildly by a few percentages.
Dissipative gas run does not differ much to the adiabatic run in dark matter variance, 
but renders significantly different $S_n$ parameters of dark matter,  bringing about more than
$10\%$ enhancement to $S_3$  at $0.1h^{-1}$Mpc and $z=0$ and being even larger at higher redshift.
Distribution patterns of gas in two hydrodynamical simulations are prominently 
different. Variance of gas at $z=0$ decreases by $\sim 30\%$
in adiabatic simulation while by $\sim 60\%$ in non-adiabatic simulation at $0.1h^{-1}$Mpc, 
the attenuation is weaker at larger scales but still obvious at $\sim 10h^{-1}$Mpc. 
$S_n$  parameters of gas are biased upward at scales $< \sim 4h^{-1}$Mpc, 
dissipative processes  give $\sim 84\%$ promotion at $z=0$ to $S_3$ at $0.1h^{-1}$Mpc 
in contrast with the $\sim 7\%$ change in adiabatic run.
The clustering segregation between gas and dark matter could have intricate implication on
modeling galaxy distribution and relevant cosmological application demanding fine
details of matter distribution in strongly nonlinear regime.
}

\authorrunning{Xiaojun Zhu \& Jun Pan }    
\titlerunning{influence of baryons on matter clustering}  
\maketitle

%
%
\section{Introduction}           
The present clustering pattern of large scale structures on cosmological scales
is generally interpreted as the growth of primordial density fluctuations mainly through
gravitation instability of the dark matter which dominates the matter content of
the Universe. While at large scales the gravitation monodrama of dark matter
is much appreciated, at small scales ignoring non-gravitational effects associated 
with galaxy formation would induce considerable systematics to relevant application. 
Quantification of such impact
of baryons, including acting scale range and strength, is strongly desired to meet the 
accuracy budget of cosmological parameter 
estimation \citep[e.g][]{ShawEtal2010, SemboloniEtal2011} and 
structure formation and evolution model 
refinement \citep[e.g.][]{StanekEtal2009, DolagEtal2009}. 
In contrast to the simplicity of 
gravitational force, physical processes baryons involved in, such as
radiative cooling, and star formation {\it etc.}, are usually very complicated 
and highly entangled, even worse is that
baryon physics works normally in strongly nonlinear regime where gravitational evolution
is already analytically intractable. 
Advanced computational facilities and algorithms, in together with accumulated 
knowledge summarized from modern observation, has enabled high resolution
hydrodynamic simulations with various treatment prescriptions for baryon physics 
plugged in \citep[e.g.][]{Teyssier2002, Springel2005}. So to date investigation on effects of 
different baryon physical processes
is carried on mainly with numerical simulations \citep[e.g.][]{vanDaalenEtal2011} though there are still a 
long way to go to build a trustworthy
machinery to capture the messy gas physics in full details.

Recent analysis of simulation data sets has shown that baryonic physical processes could alter
the matter power spectrum at $k>10h{\rm Mpc}^{-1}$ and then the weak lensing power spectrum consequently 
by some non-negligible percentage \citep{JingEtal2006}, which is confirmed and much extended in later works
\citep[e.g.][]{RuddEtal2008, HearinZentner2009, vanDaalenEtal2011, SemboloniEtal2011, CasariniEtal2012}.
The modulation to power spectrum at such small scales, speaking in terminology of halo 
mode for matter clustering, is mainly happened upon the
one-halo term  which is determined just by
mass distribution inside halos and halo mass distribution function \citep{CooraySheth2002}. 
Presence of gas in simulations does slightly boost concentration parameter
of halo mass profile \citep{LinEtal2006} and regulate the high mass branch of halo mass function measurably
\citep{StanekEtal2009, CuiEtal2011}, even with adiabatic process alone.
Such changes in halo properties actually can bring about significant amendment to strong lensing 
statistics \citep{WambsganssEtal2008} and the thermal and kinetic Sunyaev-Zel'dovich power 
spectrua \citep{BattagliaEtal2010, ShawEtal2010} as well. 

The basic scenario here is that
baryons are directly redistributed by adiabatic contraction, radiative cooling,
various feedbacks from galaxies and their central black holes, and star formation activities {\it etc.}, 
and then dark matter distribution is modified through the gravity coupling with baryons. 
Since dark matter and baryons experience different actions, it is normal to expect that their
clustering would differ with the pattern shown in the dark matter only case in complex way.
In the work of \citet{JingEtal2006} it is found that the clustering of the gas is suppressed
while that of dark matter is boosted at scales $k >1 h {\rm Mpc}^{-1}$, resulting in 
to the clustering of total matter suppress
at the level $1\%$ at $1<k<10h{\rm Mpc}^{-1}$ but then boost up to $2\%$ in the 
nonradiative run and $10\%$ in the run with star formation at $k\approx 20h{\rm Mpc}^{-1}$.
The extensive research by \citet{vanDaalenEtal2011} with AGN feedback provided a quantitative
different description, due to their different cooking recipes for gas physics, though it is still qualitatively
in agreement with \citet{JingEtal2006}. They
discovered that the $1\%$ level decreasement of power spectrum
of total matter at $z=0$ to that of pure dark matter simulation can be as low as $k\sim 0.3h {\rm Mpc}^{-1}$,
$10\%$ dropping appears at $k\sim 10h {\rm Mpc}^{-1}$, clustering enhancement
is observed somehow at $k>\sim 70h {\rm Mpc}^{-1}$. Compared with the power spectrum of pure dark matter run, 
gas in hydrodynamic simulations exhibits much less power for $k>1h{\rm Mpc}^{-1}$ while
dark matter component shows power boost at $k>10h{\rm Mpc}^{-1}$. 

The baryon influence on matter clustering and the baryon-dark matter segregation in clustering
can be better inspected with higher order correlation functions which is known to be able to
reveal more subtle details of clustering than two-point statistics like the power spectrum.
\citet{GuilletEtal2010} measured skewness of the 
MareNostrum simulation\footnote{http://astro.ft.unam.es/$\sim$marenostrum}
and also a set of dark matter only simulation. They found that relative to the matter distribution
in pure dark matter simulation, in hydrodynamic simulation the dark 
matter component has skewness decreased mildly between $0.3<r<1h^{-1}$Mpc and then boosted apparently at smaller
scales. They demonstrated that by adding an exponential gaseous disk profile to the halo model
could roughly reproduce their measurements. 

In this paper we perform count-in-cell measurements of N-body/SPH simulations in together with
a pure dark matter simulation as reference, in order to better depict the impact of gas physics on
matter distribution at higher orders in complement to
works based on power spectrum, and which also serves as an independent check to the
results of \citet{GuilletEtal2010}. Furthermore, our interests is particularly on the gas-dark matter segregation
phenomenon, i.e. distribution differences among matter of different species, 
which might cast light on the origin of the galaxy biasing. The layout
of the paper is as following, section 2 contains description
to the simulation data and estimation method for higher order correlation function, results and
their analysis are in Section 3, the last section is of summary and discussion.

\section{Count-in-cell measurements of simulation data}
\subsection{The Simulation data sets}
The simulations data sets we use are the same as in \citet{JingEtal2006}, 
which consist of three simulations produced by the GADGET2 code \citep{Springel2005}, one pure
dark simulation, one hydrodynamic simulation with adiabatic process alone, and one
hydrodynamic simulation incorporated with radiative cooling, star formation, and supernovae feedback
{\it etc.}. The three simulations are run with $512^3$ particles for each component of 
dark matter and gas, starting at $z_{ini}=120$ with the same initial condition in a cubic box of
$100h^{-1}$Mpc, and their cosmology parameters are set to 
($\Omega_m,\Omega_\Lambda,\Omega_b,\sigma_8,
n,h$)=($0.268, 0.732, 0.044, 0.85, 1, 0.71$. More details of simulations can be found
in \citet{JingEtal2006} and \citet{LinEtal2006}. Here three snapshots
at $z=0, 0.526, 1.442$ are picked up for analysis.

\subsection{Count-in-cell and higher order correlation function}
Here higher order correlation functions are those volume averaged ones, i.e.
higher order connected moments of smoothed density fluctuation fields $\delta$ by 
certain window function $w$, 
$\overline{\xi}_n=\langle \delta^n \rangle_c$. $\overline{\xi}_n$ can be estimated
through the count probability distribution function $P_N(R)$ by 
count-in-cell method. Given cubic cell of side size $R$, 
$P_N$ at this scale is the probability that a randomly thrown
cell in the catalog contains $N$ galaxies,
\begin{equation}
  P_N = \frac{1}{C} \sum_{i=1}^{C} \delta_D(N_i = N) \ .
\end{equation}
Under the usual local Poisson approximation, $P_N$ is the probability distribution
function $p(\delta)$of smoothed density fluctuation convolved with a Poisson 
kernel \citep[see the review of][]{BernardeauEtal2002}
\begin{equation}
P_N=\int_{-1}^{+\infty} p(\delta) \frac{[\langle N \rangle(1+\delta)]^N 
e^{-\langle N \rangle (1+\delta)}}{N!} d\delta \ ,
\end{equation}
where $\langle N \rangle$ is the mean count-in-cell.
Higher order correlation functions $\overline{\xi}_n$ are often expressed by the
higher order cumulants hierarchy $S_n$,
\begin{equation}
    S_n = \frac{\overline{\xi}_n}{\overline{\xi}_2^{n-1}}\  ,
\end{equation}
which can be derived from the following recursion relation
\begin{equation}
    S_n = \frac{\overline{\xi}_2 F_n}{N_c^n} - \frac{1}{n}\sum_{k=1}^{n-1}\binom{n}{k}\frac{(n-k)S_{n-k}F_k}{N_c^k}
\end{equation}
where $N_c=\langle N \rangle \overline\xi_2$ and the factorial moments
\begin{equation}
    F_k = \sum P_N(R) \times (N)_k=\langle N(N-1)\ldots (N-k+1)\rangle \ .
\end{equation} 
Explicit estimators for the variance, skewness and kurtosis are just
\begin{equation}
\begin{aligned}
\overline{\xi}_2 & =\frac{F_2}{F_1^2}-1\\
   S_3 &= \frac{F_1(F_3 - 3F_1F_2 + 2F^3_1)}{(F_2 - F^2_1)^2} \\
   S_4 &= \frac{F^2_1(F_4 - 4F_3F_1 - 3F^2_2 +12F_2F^2_1 - 6F^4_1)}{(F_2 - F^2_1)^3} \ .
\end{aligned}
\end{equation}

$P_N$ is calculated with the over-sampling algorithm to reach a $\sim 10^7$ sampling 
rate \citep{Szapudi1998a}. Probing scale $R$ is limited within 
$(0.1, 10)h^{-1}{\rm Mpc}$, the small scale cut is chosen so to ensure robust recovery of 
statistics over discreteness (normally $\langle N \rangle >0.1$ is sufficient), and
large scale limit comes from one tenth of the box size above which
correlation functions are no longer reliable by practical experiences. 
Since all simulations are evolved from the same initial condition, in the same volume and of
the same resolution, there is no need to calculate their cosmic variance (or error bars) 
if we are simply interested in their differences.

\section{influence of baryonic physics on clustering}
\subsection{clustering of dark matter}

\begin{figure}
\resizebox{\hsize}{!}{
\includegraphics{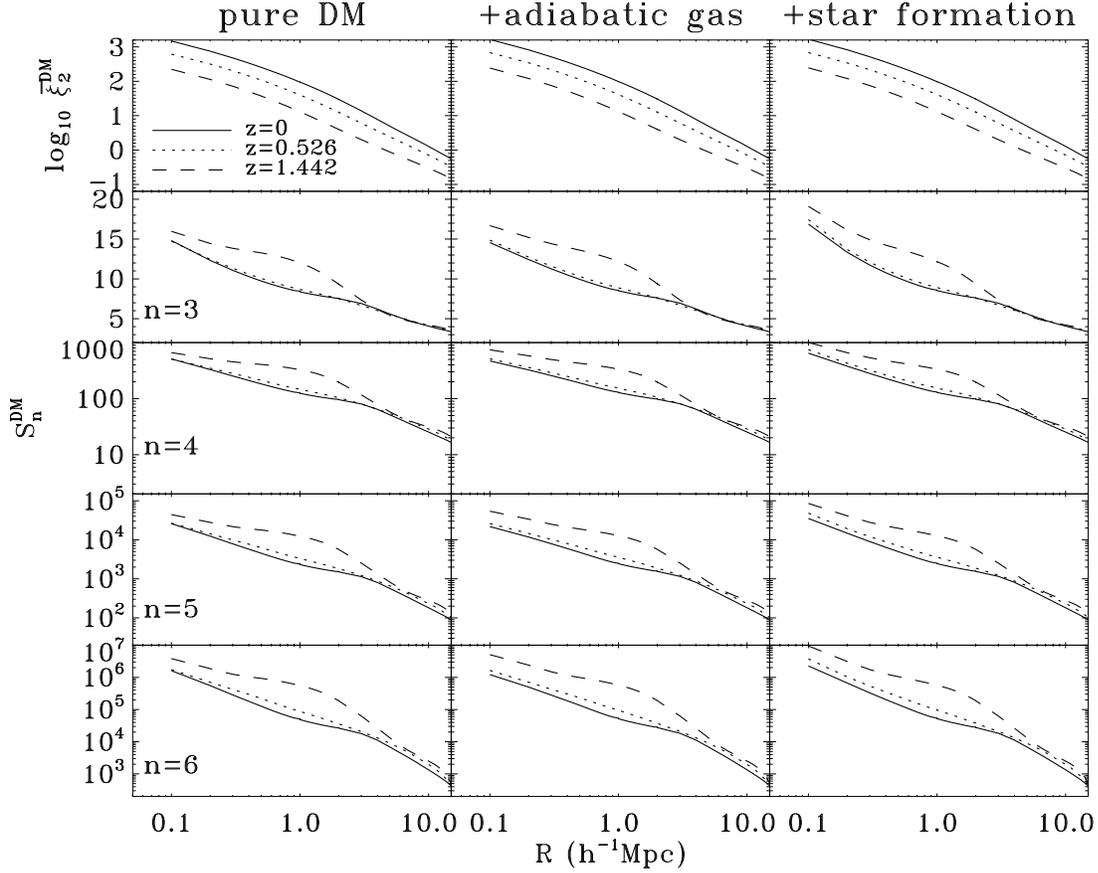}}

\caption{$\overline{\xi}_2$ and $S_n=\overline{\xi}_n/\overline{\xi}_2^{n-1}$ up 
to $n=6$ of dark matter in simulation runs of
pure dark matter (left column), with adiabatic gas physics only (middle column),  with star 
formation and other gas physics (right column). Solid, dotted and dashed
lines are of $z=0, 0.526, 1.442$ respectively.
}
\label{fig:dmSn}
\end{figure}

Correlation functions of dark matter in three simulation runs 
are illustrated in Figure~\ref{fig:dmSn}, showing analogous redshift
evolution history. It has been checked that if we rescale $\overline{\xi}_2$ with
$D(z)$ the growth rate of large scale 
structure \citep[see][for approximate formula]{LahavEtal1991}, variances at different redshift
are in good agreement at scales where $\overline{\xi}_2(R,z=0) < 1$,  at smaller scales
$\overline{\xi}_2(z)D^2(z=0)/D^2(z)$ becomes lower with increasing $z$ as what is well known.

$S_n$ parameters are apparently
larger at higher redshift at scales $R< \sim 3h^{-1}$Mpc, 
the decrement from $z=1.442$ to
$z=0.526$ in $S_n$ parameters is much bigger than that from $z=0.526$ to $z=0$. 
Recall that $\overline{\xi}_{n>2}$s are measures of
non-Gaussianity and our simulations are evolved from Gaussian initial condition, intuitively
non-Gaussianity would increase when redshift decreases, in
line with the growth of gravitational nonlinearity. Let the cosmic scale factor $a=1/(1+z)$ and
$\Delta a >0$ be a small increment to $a$, 
as $S_n(a+\Delta a)<S_n(a)$, $S_n=\overline{\xi}_n/\overline{\xi}_2^{n-1}$,
$\overline{\xi}_n(a+\Delta a) > \overline{\xi}_n(a)$ and
at small scales $\overline{\xi}_n>0$, there is
\begin{equation}
\log \overline{\xi}_n(a+\Delta a)-\log \overline{\xi}_n(a) < (n-1)
\left[ \log\overline{\xi}_2(a+\Delta a) - \log \overline{\xi}_2(a) \right] \ ,
\end{equation}
subsequently
\begin{equation}
\frac{d\log \overline{\xi}_n}{d\log a} < (n-1)\frac{d\log \overline{\xi}_2}{d\log a}\ , \ \ \ for\ n>2\ ,
\end{equation}
which establishes an interesting relation between the evolution rates of higher order correlation functions
and that of the two point correlation function in strongly nonlinear regime. 

At scales $R>\sim 3h^{-1}$Mpc $S_n$s show little redshift dependence, there is
$S_n(a<1)\approx S_n(a=1)$ as perturbation theory predicted \citep{FosalbaGaztanaga1998}.
However the approximation is not perfect  in our raw results, at large scales there are differences at
different epochs at level of several percentages, $S_n$s at higher redshift turns to be slightly larger. 
It is known that there is blemish rooted in the Zel'dovich approximation based
initial condition generator,  the resulting correction to measured $S_n$s at leading order decays with redshift
at rate roughly $\propto [D(z)/D(z_{ini})]^{-1}$. But the effect of the bias is that true $S_n$s are larger than 
measured values \citep{Scoccimarro1998, FosalbaGaztanaga1998}, so that the offsets between different redshift will be
even higher than what is shown in Figure~\ref{fig:dmSn}. The puzzle could be dynamical rather than systematical biases, here we just leave the issue to future
work.

\begin{figure}
\resizebox{\hsize}{!}{\includegraphics{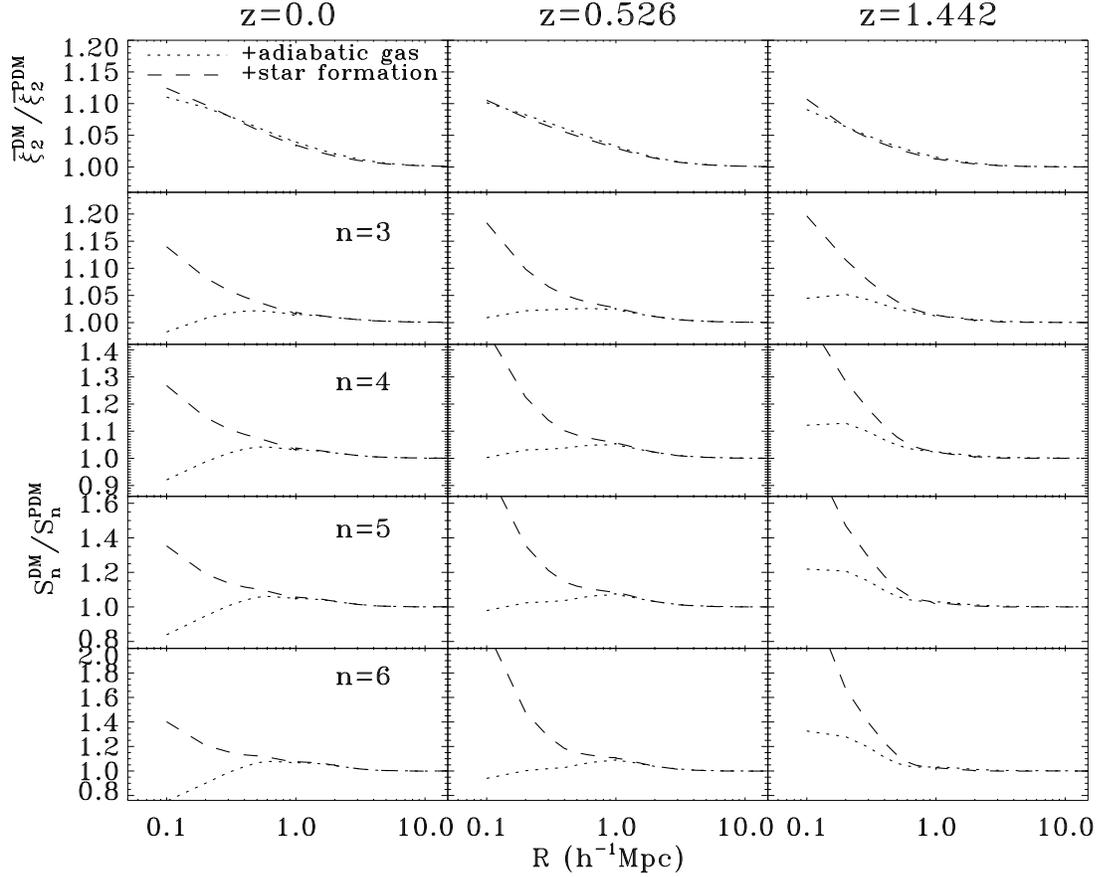}}
\caption{Influence of gas physics on dark matter distribution: 
ratios of correlation functions of dark matter in hydrodynamical
simulations to those of dark matter in pure dark matter simulation.}
\label{fig:dmSnratio}
\end{figure}

Influence of gas physics on dark matter distribution is displayed in Figure~\ref{fig:dmSnratio}.
Enhancement to $\overline{\xi}_2$ induced by gas is mild and 
increases to $\sim 10\%$ at $\sim 0.1h^{-1}$Mpc, which
is consistent with previous works \citep[e.g.][]{JingEtal2006, GuilletEtal2010}. Gas physics other than 
adiabatic process does not bring significant extra modulation to
the two point correlation function of dark matter, their effects are seen in higher order functions.

From Figure~\ref{fig:dmSnratio} it is clear that dark matter clustering is immune to baryons at 
large scales, stage that gas physics play is mainly on scales $R<\sim 1h^{-1}$Mpc.
In the adiabatic run, at $z=1.442$, $S_n^{DM}/S_n^{PDM}$ only becomes larger than one
on scales smaller than $1h^{-1}$Mpc, it is smaller at $z=0.526$, then the
boost switches to suppression at scales $<0.2h^{-1}$Mpc. The level of impact
is not very large, at $0.1h^{-1}$Mpc skewness $S_3$ is 
increased by $\sim 5\%$ at $z=1.442$ and then decreased by $\sim 2\%$ at $z=0$, variation
in kurtosis at such scale is $\sim +12\%$ at $z=1.442$ but $\sim -8\%$ at $z=0$. 

Effects of star formation activities and other gas physics are much stronger than
the adiabatic process, amplitudes of $S_n$s at all redshift are raised significantly, however
the increment decreases at lower redshift, for instance the relative enhancement to
$S_3$ and $S_4$ at $R=0.1h^{-1}$Mpc is $\sim 20\%$ and $\sim 52\%$ at $z=1.442$ 
but drops down to $\sim 14\%$ and $\sim 27\%$ at $z=0$ respectively.
It appears that if we are about to investigate differences among models built with
different baryonic processes through clustering analysis of dark matter, we should
concentrate on higher order statistics on sub-mega parsec scales, and preferentially
at higher redshift.

\subsection{clustering of baryonic gas}

\begin{figure}
\resizebox{\hsize}{!}{\includegraphics{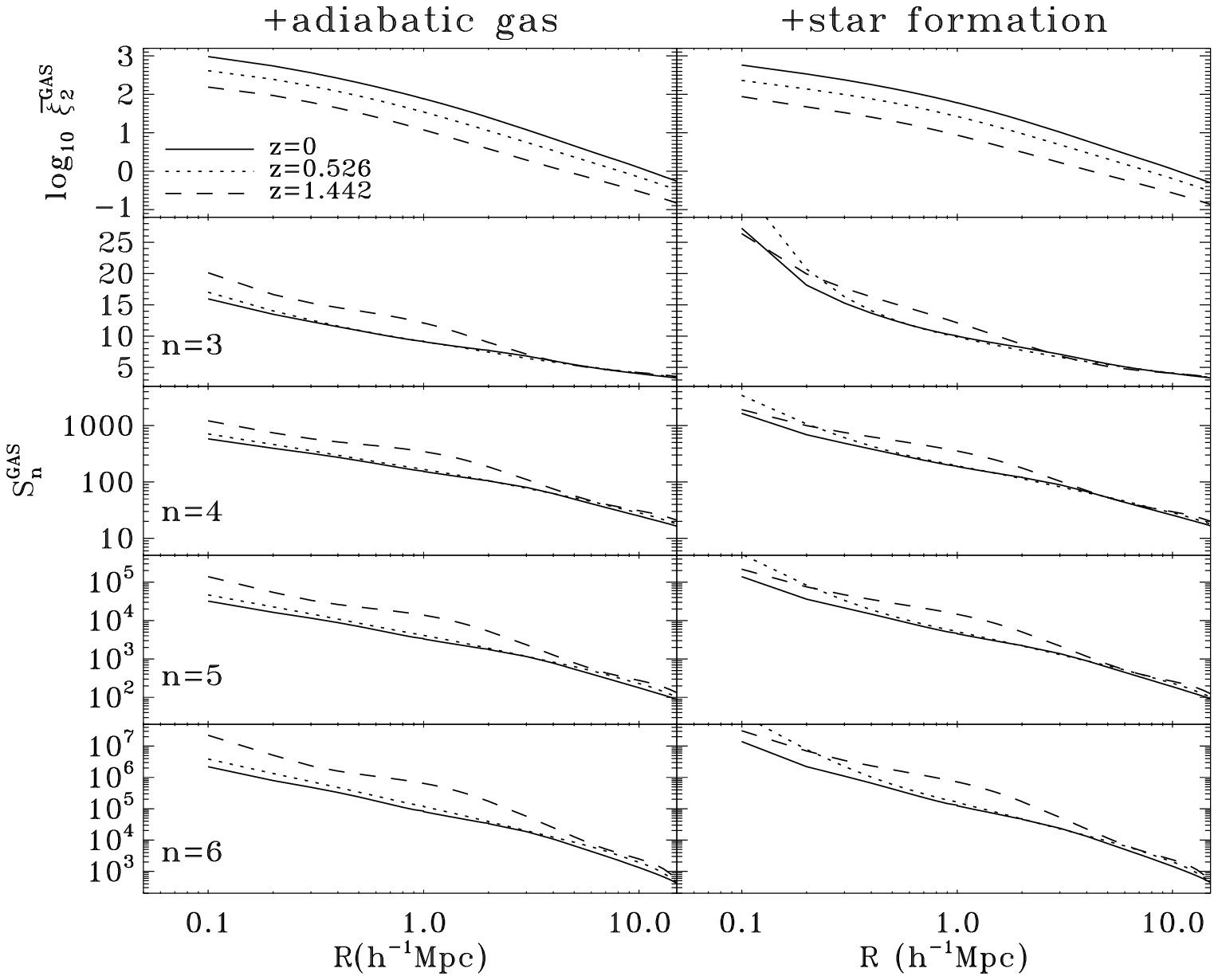}}
\caption{$\overline{\xi}_2$ and $S_n$ of gas in hydrodynamic simulations.}
\label{fig:gasSn}
\end{figure}

Correlation functions of gas are presented in Figure~\ref{fig:gasSn}.
Redshift evolution of correlation functions of gas in the adiabatic simulation is
similar to the dark matter component as in Figure~\ref{fig:dmSn}, 
$S_n$ being larger at higher redshift at scales less than $\sim 3h^{-1}$Mpc. 
There is more complexity in the non-adiabatic
hydrodynamic simulation, $S_n$ parameters demonstrate intricate evolution
path at scales $<\sim 0.2h^{-1}$Mpc, which is probably reflection of the composite
action from competing gas physical processes, e.g. radiative cooling versus feedbacks
from supernovae. To separate effect of individual
ingredient of gas physics, series simulations installed with different prescriptions
are definitely needed, like what \citet{vanDaalenEtal2011} did.

\begin{figure}
\resizebox{\hsize}{!}{\includegraphics{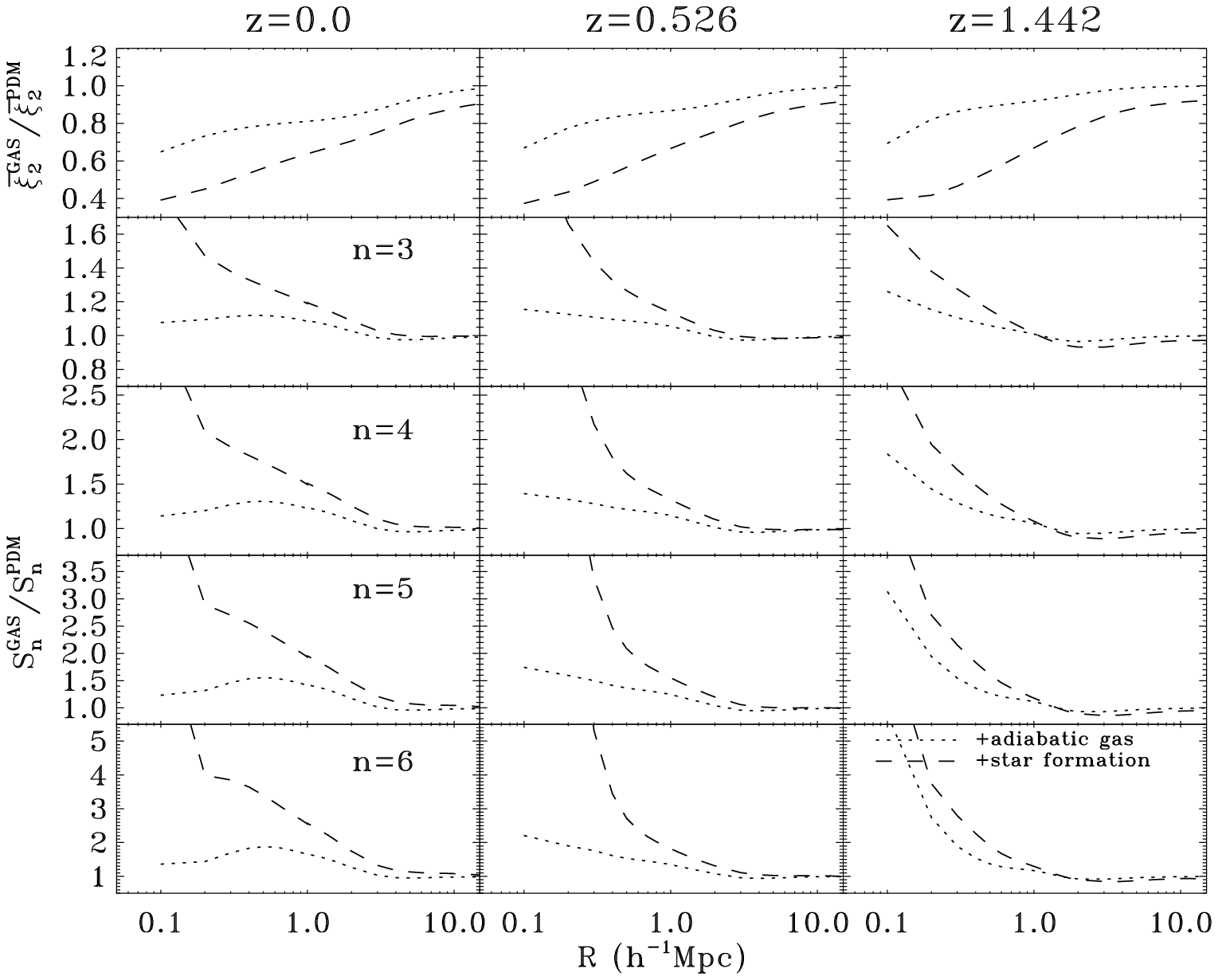}}
\caption{Comparison of correlation functions of gas in hydrodynamic simulations
to those of dark matter in the pure dark matter run.}
\label{fig:gas2pdm}
\end{figure}

In hydrodynamical simulations, gas distribution deviate from the dark matter distribution in 
pure dark matter more obviously than their dark matter counterpart (Figure~\ref{fig:gas2pdm}), 
in aspects of both amplitudes of correlation functions and affected scale range, while again
the effects become weaker at lower redshift. All $S_n$s of gas are boosted significantly at
scales $<\sim 4 h^{-1}$Mpc, 
in agreement with \citet{JingEtal2006} variance of gas $\overline{\xi}_2$
is depressed severely on fairly broad scales extending to around $10h^{-1}$Mpc, 
at $z=0$ $\overline{\xi}_2$ decreases by $\sim 30\%$ in adiabatic simulation while 
by $\sim 60\%$ in non-adiabatic simulation at $0.1h^{-1}$Mpc. 
Bifurcation due to differences in 
gas physical mechanisms employed in simulations is observed too, non-adiabatic gas physics
inducing stronger variation to the clustering of matter, 
there is $\sim 84\%$ and $\sim 220\%$ promotion at $z=0$ to $S_3$ and
$S_4$ respectively at $0.1h^{-1}$Mpc but only moderate 
$\sim 7\%$ and $\sim 14\%$ fortification in adiabatic simulation.

\subsection{gas-dark matter clustering segregation}

\begin{figure}
\resizebox{\hsize}{!}{\includegraphics{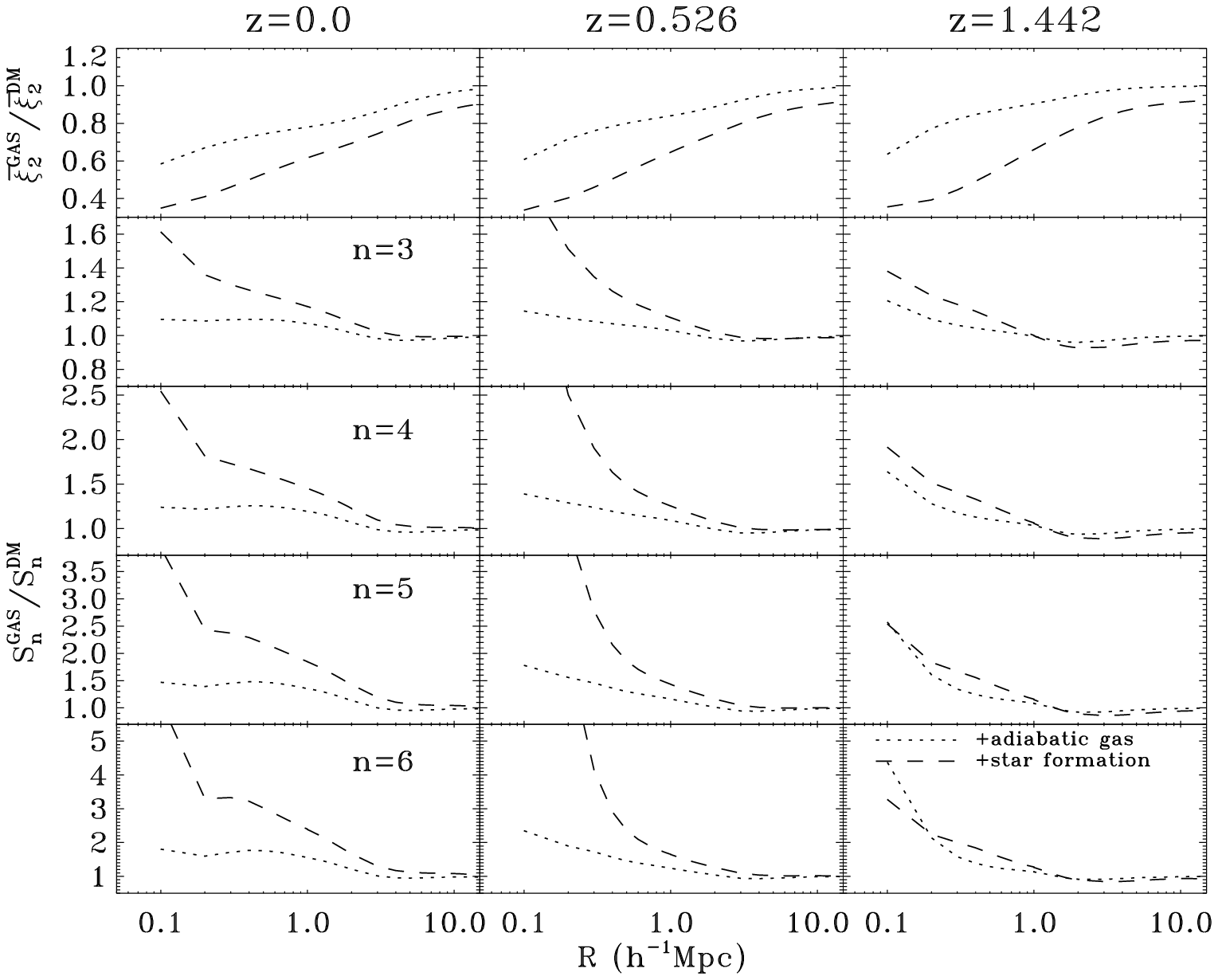}}
\caption{Differences between distributions of gas and dark matter in hydrodynamic
simulations.}
\label{fig:gas2dm}
\end{figure}

It is clear now that including gas can result in distribution patterns of both dark matter
and gas different to the case of pure dark matter at very small scales. From results
shown above it appears that gas distribution is more severely affected than dark matter, which
is easy to understand as all gas physics directly act on gas, whilst dark matter is influenced
only through its gravitational coupling to the baryons. Direct comparison of
correlation functions of gas with dark matter is in Figure~\ref{fig:gas2dm}, 
we can see that for two point correlation function differences appear at scales
as large as $\sim 10h^{-1}$Mpc but for for higher order cumulants the departure
scale is $\sim 4h^{-1}$Mpc.
Volume averaged two point correlation
function of gas is obviously lower than the dark matter, at $0.1h^{-1}$Mpc the gap
could be as large as $\sim 40-60\%$, in adiabatic run the difference at $z=0$ is around $20\%$
already at $1h^{-1}$Mpc and in the non-adiabatic run it becomes $\sim 40\%$ at such scale.
However $S_n$ parameters of gas are much larger than dark matter, for example
at $0.1h^{-1}$Mpc enhancement to the skewness $S_3$
at $z=0$ is about $10\%$ in adiabatic run and $\sim 62\%$ in the non-adiabatic run.
The results indicate that distribution of gas has a longer tail than that of dark matter,
in another words, there are more highly concentrated small clumps of gas than
dark matter. 

Biasing of gas to dark matter in $S_n$s at $z=0.526$ is largest in both hydrodynamic
simulations. We conjecture that effects of gas physics must reach summit at some redshift 
$z>0$ and then relax after that, entering a more passive
evolution stage. If galaxy assembly rate is strongly correlated with accumulated
effect of gas physics, likely there would be an apex within that time interval, of course the exact 
peak time may depends on how baryon processes are cooked.

\section{summary and discussion}
In this paper correlation functions 
upto the sixth order are estimated from count-in-cell analysis of
dark matter and gas respectively in three
simulations, the pure dark matter run, the run with adiabatic gas process, and
the one with star formation activities and other gas physics.
Major results about influence on matter clustering are in the following.
\begin{enumerate}[(1)]
\item Compared with the case of
pure dark matter, baryon physical processes introduce non-negligible modulation to 
clustering of dark matter, affected regime for dark matter is
at scales less than $1h^{-1}$Mpc. 
Adiabatic process alone strengthens $\overline{\xi}_2$
by $\sim 10\%$ at scale $0.1h^{-1}$Mpc, which is insensitive to redshift; $S_n$ parameters
in the run deviate from pure dark matter results rather mildly, at $0.1h^{-1}$Mpc
skewness $S_3$ evolves from $\sim 5\%$ lifting at $z=1.442$ to $\sim 2\%$ falling
at $z=0$, meanwhile the difference in kurtosis $S_4$ changes from
$\sim +12\%$ to negative $\sim -8\%$. In the run with dissipative gas processes
$\overline{\xi}_2$ does not differ much to the adiabatic run, 
but $S_n$ parameters all are increased significantly, bringing 
$\sim +14\%$ to $S_3$ and $\sim +27\%$ to $S_4$ at $0.1h^{-1}$Mpc and $z=0$, and
the amplitude of change is larger in higher redshift.
\item Gas distribution in hydrodynamic simulations is much more strongly modified than
dark matter component. Two point correlation function of gas at $z=0$ decreases by $\sim 30\%$
in adiabatic simulation while by $\sim 60\%$ in non-adiabatic simulation at $0.1h^{-1}$Mpc, 
the attenuation is weaker at larger scales but still obvious at $\sim 10h^{-1}$Mpc. 
$S_n$  parameters of gas are biased upward at scales $< \sim 4h^{-1}$Mpc, 
dissipative processes add prominently more power to them, 
giving a $\sim 84\%$ promotion at $z=0$ to $S_3$ at $0.1h^{-1}$Mpc against the moderate 
$\sim 7\%$ fortification in adiabatic simulation.
\item There is clustering segregation between gas and dark matter in the same simulation. $\overline{\xi}_2$ 
of gas is already lower than dark matter counterpart at $\sim 10h^{-1}$Mpc, which is down at $0.1h^{-1}$Mpc
by $\sim40\%$ and $\sim62\%$ in the adiabatic run and the non-adiabatic run respectively. $S_n$s 
of gas are larger than dark matter at scales $<4h^{-1}$Mpc, $S_3$ of gas in adiabatic run is leveled up
by $\sim 10\%$ while by $\sim 60\%$ in non-adiabatic run at $0.1h^{-1}$Mpc.
Biasing of gas to dark matter is much stronger in non-adiabatic simulation than the adiabatic only run, and
the maximal bias is achieved at certain redshift $z>0$.
\end{enumerate}

It is shown in this work that difference in distribution of dark matter originated from various mechanisms of 
gas physics can not be effectively distinguished at the second order level, though apparent
discrepancy appear in gas. It would benefit those applications which rely on second order statistical 
properties of dark matter only, but once going to higher orders one has to consider the systematics 
brought forward by gas. 

Biasing of gas to dark matter is a more interesting problem, aside from that it may be a serious challenge to precision 
cosmology such as the modeling to Sunyaev-Zeldovich effects \citep[e.g.][]{ShawEtal2010, BattagliaEtal2010}. 
We know that galaxies are biased tracers of dark matter distribution, but galaxies
are in fact products of gas physics, probably it is more reasonable to assume
that galaxies are actually tracing gas instead of the dark matter. We conjecture that by
the decomposition stochasticity and nonlinearity of galaxy bias would be greatly reduced.
Standard methods exploring relation between galaxies and their host halos, such as the
halo occupation distribution model \citep{BerlindWeinberg2002} and 
the conditional luminosity function model \citep{YangEtal2003}, generally use the
two point correlation function summarized from pure dark matter simulation as reference
to the measured galaxy two point correlation function. Data points of galaxy two point correlation 
function usually are at scales from $\sim 0.1h^{-1}$Mpc to a few mega parsec within which
unfortunately the matter distribution underlying to galaxies is not the same as what is
in pure dark matter universe, we might have to quantify the amplitude of this kind of systematical
bias before presenting estimation of number of a particular type of galaxies in halos.

\begin{acknowledgements}
This work is supported by the NSFC through
grants of Nos. 10873035, and 11133003.
JP acknowledges the One-Hundred-Talent fellowship of CAS. 
We thank Weipeng Lin for his kindness of providing his N-body
simulation data,  the simulations were done at Shanghai Supercomputer 
Center by the supports of Chinese National 863 project (No. 2006AA01A125).

\end{acknowledgements}


\begin{thebibliography}{24}
\expandafter\ifx\csname natexlab\endcsname\relax\def\natexlab#1{#1}\fi

\bibitem[{{Battaglia} {et~al.}(2010){Battaglia}, {Bond}, {Pfrommer}, {Sievers},
  \& {Sijacki}}]{BattagliaEtal2010}
{Battaglia}, N., {Bond}, J.~R., {Pfrommer}, C., {Sievers}, J.~L., \& {Sijacki},
  D. 2010, \apj, 725, 91

\bibitem[{{Berlind} \& {Weinberg}(2002)}]{BerlindWeinberg2002}
{Berlind}, A.~A., \& {Weinberg}, D.~H. 2002, \apj, 575, 587

\bibitem[{{Bernardeau} {et~al.}(2002){Bernardeau}, {Colombi}, {Gazta{\~n}aga},
  \& {Scoccimarro}}]{BernardeauEtal2002}
{Bernardeau}, F., {Colombi}, S., {Gazta{\~n}aga}, E., \& {Scoccimarro}, R.
  2002, \physrep, 367, 1

\bibitem[{{Casarini} {et~al.}(2012){Casarini}, {Bonometto}, {Borgani}, {Dolag},
  {Murante}, {Mezzetti}, {Tornatore}, \& {La Vacca}}]{CasariniEtal2012}
{Casarini}, L., {Bonometto}, S.~A., {Borgani}, S., {Dolag}, K., {Murante}, G.,
  {Mezzetti}, M., {Tornatore}, L., \& {La Vacca}, G. 2012, ArXiv e-prints,
  astro-ph/1203.5251

\bibitem[{{Cooray} \& {Sheth}(2002)}]{CooraySheth2002}
{Cooray}, A., \& {Sheth}, R. 2002, \physrep, 372, 1

\bibitem[{{Cui} {et~al.}(2011){Cui}, {Borgani}, {Dolag}, {Murante}, \&
  {Tornatore}}]{CuiEtal2011}
{Cui}, W., {Borgani}, S., {Dolag}, K., {Murante}, G., \& {Tornatore}, L. 2011,
  ArXiv e-prints, astro-ph/1111.3066

\bibitem[{{Dolag} {et~al.}(2009){Dolag}, {Borgani}, {Murante}, \&
  {Springel}}]{DolagEtal2009}
{Dolag}, K., {Borgani}, S., {Murante}, G., \& {Springel}, V. 2009, \mnras, 399,
  497

\bibitem[{{Fosalba} \& {Gazta{\~n}aga}(1998)}]{FosalbaGaztanaga1998}
{Fosalba}, P., \& {Gazta{\~n}aga}, E. 1998, \mnras, 301, 503

\bibitem[{{Guillet} {et~al.}(2010){Guillet}, {Teyssier}, \&
  {Colombi}}]{GuilletEtal2010}
{Guillet}, T., {Teyssier}, R., \& {Colombi}, S. 2010, \mnras, 405, 525

\bibitem[{{Hearin} \& {Zentner}(2009)}]{HearinZentner2009}
{Hearin}, A.~P., \& {Zentner}, A.~R. 2009, \jcap, 4, 32

\bibitem[{{Jing} {et~al.}(2006){Jing}, {Zhang}, {Lin}, {Gao}, \&
  {Springel}}]{JingEtal2006}
{Jing}, Y.~P., {Zhang}, P., {Lin}, W.~P., {Gao}, L., \& {Springel}, V. 2006,
  \apjl, 640, L119

\bibitem[{{Lahav} {et~al.}(1991){Lahav}, {Lilje}, {Primack}, \&
  {Rees}}]{LahavEtal1991}
{Lahav}, O., {Lilje}, P.~B., {Primack}, J.~R., \& {Rees}, M.~J. 1991, \mnras,
  251, 128

\bibitem[{{Lin} {et~al.}(2006){Lin}, {Jing}, {Mao}, {Gao}, \&
  {McCarthy}}]{LinEtal2006}
{Lin}, W.~P., {Jing}, Y.~P., {Mao}, S., {Gao}, L., \& {McCarthy}, I.~G. 2006,
  \apj, 651, 636

\bibitem[{{Rudd} {et~al.}(2008){Rudd}, {Zentner}, \& {Kravtsov}}]{RuddEtal2008}
{Rudd}, D.~H., {Zentner}, A.~R., \& {Kravtsov}, A.~V. 2008, \apj, 672, 19

\bibitem[{{Scoccimarro}(1998)}]{Scoccimarro1998}
{Scoccimarro}, R. 1998, \mnras, 299, 1097

\bibitem[{{Semboloni} {et~al.}(2011){Semboloni}, {Hoekstra}, {Schaye}, {van
  Daalen}, \& {McCarthy}}]{SemboloniEtal2011}
{Semboloni}, E., {Hoekstra}, H., {Schaye}, J., {van Daalen}, M.~P., \&
  {McCarthy}, I.~G. 2011, \mnras, 417, 2020

\bibitem[{{Shaw} {et~al.}(2010){Shaw}, {Nagai}, {Bhattacharya}, \&
  {Lau}}]{ShawEtal2010}
{Shaw}, L.~D., {Nagai}, D., {Bhattacharya}, S., \& {Lau}, E.~T. 2010, \apj,
  725, 1452

\bibitem[{{Springel}(2005)}]{Springel2005}
{Springel}, V. 2005, \mnras, 364, 1105

\bibitem[{{Stanek} {et~al.}(2009){Stanek}, {Rudd}, \&
  {Evrard}}]{StanekEtal2009}
{Stanek}, R., {Rudd}, D., \& {Evrard}, A.~E. 2009, \mnras, 394, L11

\bibitem[{{Szapudi}(1998)}]{Szapudi1998a}
{Szapudi}, I. 1998, \apj, 497, 16

\bibitem[{{Teyssier}(2002)}]{Teyssier2002}
{Teyssier}, R. 2002, \aap, 385, 337

\bibitem[{{van Daalen} {et~al.}(2011){van Daalen}, {Schaye}, {Booth}, \& {Dalla
  Vecchia}}]{vanDaalenEtal2011}
{van Daalen}, M.~P., {Schaye}, J., {Booth}, C.~M., \& {Dalla Vecchia}, C. 2011,
  \mnras, 415, 3649

\bibitem[{{Wambsganss} {et~al.}(2008){Wambsganss}, {Ostriker}, \&
  {Bode}}]{WambsganssEtal2008}
{Wambsganss}, J., {Ostriker}, J.~P., \& {Bode}, P. 2008, \apj, 676, 753

\bibitem[{{Yang} {et~al.}(2003){Yang}, {Mo}, \& {van den Bosch}}]{YangEtal2003}
{Yang}, X., {Mo}, H.~J., \& {van den Bosch}, F.~C. 2003, \mnras, 339, 1057

\end{thebibliography}
\end{document}